\documentclass[conference]{IEEEtran}
\IEEEoverridecommandlockouts

\usepackage{cite}
\usepackage{amsmath,amssymb,amsfonts}
\usepackage{algorithmic}
\usepackage{graphicx}
\usepackage{algorithm}
\usepackage{subfig}
\usepackage{textcomp}
\usepackage{xcolor}
\usepackage{booktabs}
\def\BibTeX{{\rm B\kern-.05em{\sc i\kern-.025em b}\kern-.08em
    T\kern-.1667em\lower.7ex\hbox{E}\kern-.125emX}}

\usepackage{multirow}

\begin{document}

\title{Beyond Static Thresholds:\\Adaptive RRC Signaling Storm Detection\\with Extreme Value Theory}


\author{\IEEEauthorblockN{Dang Kien Nguyen}
\IEEEauthorblockA{\textit{Standards \& Technology, Ericsson} \\
\textit{Communication Systems Department, EURECOM}\\
Massy, France \\
dang.kien.nguyen@ericsson.com}
\and
\IEEEauthorblockN{Rim El Malki}
\IEEEauthorblockA{\textit{Standards \& Technology} \\
\textit{Ericsson}\\
Massy, France \\
rim.el.malki@ericsson.com}
\and
\IEEEauthorblockN{Filippo Rebecchi}
\IEEEauthorblockA{\textit{Standards \& Technology} \\
\textit{Ericsson}\\
Massy, France \\
filippo.rebecchi@ericsson.com}
\and
\IEEEauthorblockN{Raymond Knopp}
\IEEEauthorblockA{\textit{Communication Systems Department} \\
\textit{EURECOM}\\
Sophia Antipolis, France \\
raymond.knopp@eurecom.fr}
\and
\IEEEauthorblockN{Melek Önen}
\IEEEauthorblockA{\textit{Digital Security Department} \\
\textit{EURECOM}\\
Sophia Antipolis, France \\
melek.onen@eurecom.fr}
}

\maketitle

\begin{abstract}
In 5G and beyond networks, the radio communication between a User Equipment (UE) and a base station (gNodeB or gNB), also known as the air interface, is a critical component of network access and connectivity. During the connection establishment procedure, the Radio Resource Control (RRC) layer can be vulnerable to signaling storms, which threaten the availability of the radio access control plane. These attacks may occur when one or more UEs send a large number of connection requests to the gNB, preventing new UEs from establishing connections. In this paper, we investigate the detection of such threats and propose an adaptive threshold-based detection system based on Extreme Value Theory (EVT). The proposed solution is evaluated numerically by applying simulated attack scenarios based on a realistic threat model on top of real-world RRC traffic data from an operator network. We show that, by leveraging features from the RRC layer only, the detection system can not only identify the attacks but also differentiate them from legitimate high-traffic situations. The adaptive threshold calculated using EVT ensures that the system works well under diverse threat scenarios.
The results show high accuracy, precision, and recall values (above 93\%), and a low detection latency even under complex conditions.
\end{abstract}


\section{Introduction}

With the ongoing development of 6G,  mobile networks are expected to become increasingly important across a wide spectrum of domains, from emergency communications, smart cities, traffic safety and efficiency, and more~\cite{elayoubi20165g}. With use cases continuously diversifying and becoming more demanding, including mission-critical and security-sensitive applications, future networks must be able to handle massive volumes of data traffic with high reliability, low latency, and adaptive resource management~\cite{ettiane2021toward}.


To initiate a radio connection between the User Equipment (UE) and the Radio Access Network (RAN), an establishment procedure is carried out over the air interface with the base station (gNB in 5G terminology). During this process, the Radio Resource Control (RRC) layer at the gNB plays a central role in managing the connection setup and maintenance, ensuring a reliable and controlled radio link between the UE and the network~\cite{hailu2018rrc,mukherjee20195g}.


RRC signaling storms are availability attacks that target the gNB control plane. Malicious UEs (MUEs) continuously trigger the RRC connection procedure, without ever completing it. These attacks exploit the resource reservation mechanism inherent to the RRC connection procedure, i.e., a design that has remained essentially unchanged from 3G through to the forthcoming 6G standards. Similar to a Denial-of-Service (DoS) attack, the resulting flood of open connections may lead to the exhaustion of radio and processing resources of the gNB, blocking legitimate UEs from establishing connections~\cite{tabiban2023signaling}. Detecting and mitigating these signaling storms is particularly complex because RRC establishment precedes network authentication. At that stage, the gNB lacks any trustworthy identity information about requesting UEs.

State-of-the-art RRC signaling-storm detectors usually monitor features available at higher protocol layers. This adds latency for both detection and mitigation, even though the attack originates at the lower layers of the protocol stack~\cite{abdelrahman2016detecting}. Moreover, current solutions suffer from two key shortcomings: 
\begin{itemize}
    \item They use fixed threshold values that are unsuitable for varying traffic conditions of real networks~\cite{nguyen2025rrc}. 
    \item They ignore legitimate high-load situations. During emergencies, such as natural disasters or large public events, many Benign UEs (BUEs) may simultaneously request RRC connections. A gNB must be able to differentiate such normal behavior from actual attacks~\cite{hoffmann2023signaling,feng2023research}.
\end{itemize}

\subsection{Contributions}

With the objective of providing an effective and practical detection solution to RRC signaling storms, the main contributions and novelties of this work are as follows:
\begin{itemize}
    \item We introduce an adaptive Extreme Value Theory (EVT)-based thresholding system that detects RRC signaling storms while discriminating them  from legitimate high-load situations. The threshold self-adapts to varying traffic conditions without the need for manual configuration.
    \item We validate the detector using real-world mobile traffic traces collected from operational networks, and enriched with  theoretical model for attacks/high-load situations.
    \item We carry out an extensive performance evaluation that covers a wide spectrum of traffic conditions and abnormal scenarios - from minor disruptions, to severe unavailability drops at the gNB. Our detector consistently achieves $>93\%$ accuracy, precision and recall across all scenarios and flags anomalies with an average latency of 2.72 s. The EVT focus on the tail of the event distribution yields clear advantages over conventional baseline techniques (e.g., that assumes Gaussian distribution). 
\end{itemize}


\subsection{Literature Review}
\label{Literature}
This section reviews existing work on RRC signaling storm detection,  adaptive thresholds for signaling storms, and anomaly and DoS detection.

The RRC signaling storm is implemented to understand its impact on the gNB in~\cite{nguyen2025rrc}. The authors also implement a simple detection system with RRC layer features, but use a fixed threshold, which cannot be used in different scenarios, e.g., different traffic loads, or different base stations. In~\cite{feng2023research}, the authors utilize a Machine Learning (ML) model along with an adaptive threshold to detect signaling storms in 4G/5G networks. However, they do not specifically look at RRC signaling storms at the gNB, but other signaling storms targeting the core network. Their proposed framework also does not differentiate between attacks and benign cases, and only tracks the impacted components inside the network. The work in~\cite{hoffmann2023signaling} uses a statistical method to calculate a Key Performance Indicator and then compares the value with a threshold. The authors also do not use an adaptive threshold, and their work does not distinguish attacks from legitimate high-loads. Similarly,~\cite{pavloski2019detecting} uses a fixed threshold for the number of mobile access requests to the cloud. Lastly, while two studies in~\cite{escudero2019detecting,abdelrahman2016detecting} use the Dempster-Shafer theory of evidence and Random Neural Network (RNN) for their detection methods instead of threshold-based detection systems, they cannot identify a legitimate high-load.

Extreme Value Theory is a statistical method that is suitable for adaptive threshold detection systems, and has been applied in multiple studies~\cite{lin2024hierarchical,li2021fluxev,siffer2017anomaly}. Thus, this method is chosen for the present research and will be explained in more detail in section \ref{EVT}.
In other research focused on adaptive thresholds, authors often use these thresholds alongside machine learning, where the adaptive thresholds serve as a supporting features rather than make  final decisions~\cite{yuan2024dynamic,niu2023two}.

In summary, the current state-of-the-art does not have a sufficiently adaptive detection system that can work under different conditions, especially a detection system that can identify high-load situations.

\section{Background}
\label{Background}
This section provides background information on the RRC establishment procedure in 5G, which is essential for understanding the problem, and introduces the Extreme Value Theory, which will be used for the adaptive threshold mechanism in the proposed solution.

\subsection{RRC Establishment Procedure}
\begin{figure}[htbp]
\centerline{\includegraphics[width=0.8\linewidth]{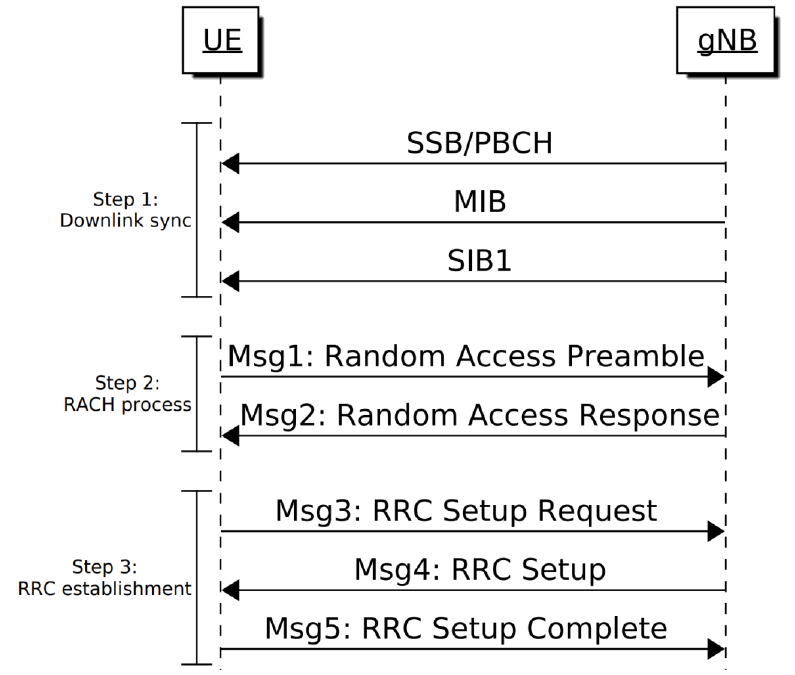}}
\caption{5G Connection Establishment~\cite{nguyen2025rrc}.}
\label{con_establishment}
\end{figure}

In the 5G connection establishment procedure, as shown in Fig.~\ref{con_establishment}, a new UE begins with the downlink synchronization and Random Access Channel (RACH) processes, in which the device receives the initial configuration to start subsequent processes.
As part of the establishment, after the RACH process, the RRC process contains three messages.
The UE starts the RRC connection establishment by sending an RRC Setup Request (Msg3), which contains the UE's identity and an establishment cause. However, the UE's identity in the message can be either an S-TMSI (SAE Temporary Mobile Subscriber Identity), which is known by the network, or a random value. Upon receiving Msg3, the gNB verifies the provided information. If the Msg3 is valid, the gNB will reserve an RRC resource for the UE, and respond with an RRC Setup (Msg4), which carries more configuration information.
To finalize the RRC procedure, the UE then sends an RRC Setup Complete (Msg5) to confirm the connection, i.e., goes from the RRC\_IDLE to the RRC\_CONNECTED state, and begins the data transmission.

\subsection{Anomaly Detection with Extreme Value Theory}
\label{EVT}

Anomaly detection is often considered an unsupervised learning task that aims to detect abnormal behaviors when they deviate significantly from normal data, without depending on the labeled data.
Common approaches to anomaly detection calculate a suitable threshold based on the distribution of normal data and compare it with new data to distinguish between normal and abnormal events. Known limitations of such an approach are that it requires prior knowledge of the distribution parameters, and struggles when there are concept drifts in the data trend~\cite{lin2024hierarchical}. When these concept drifts occur, it is difficult to estimate the necessary parameters (e.g., the mean and standard deviation of the distribution), which results in many False Positives (FPs) and False Negatives (FNs).

To solve this issue, the EVT is a branch of statistics that does not make any assumptions about the original distribution of the entire dataset~\cite{siffer2017anomaly}. In reality, data do not necessarily follow common distributions (e.g., Gaussian, uniform), and the distribution may change with time. The EVT avoids making strong hypotheses on the original distribution. This theory focuses on the tails of the distribution by finding the law of extreme events and then detecting any anomalies among them. For time series data, EVT is applied using a moving window to identify the variation of the data stream. Thus, EVT can be aware of any concept drift in the data, and still not overlook the anomalies when they occur.

Peaks Over Threshold (POT) is a method of EVT that focuses on the distribution of extreme events in the data. In this method, the Extreme Value Distribution (EVD) laws state that extreme events have the same type of distribution, and this distribution is independent of normal events. In time series data streams, the extreme events follow a Generalized Pareto Distribution (GPD)~\cite{lin2024hierarchical}. 

\textbf{\textit{Notations:}}
These notations will be used in the equations for the EVT method in this paper.

\begin{itemize}
    \item $t$: The initial threshold, or peak threshold, to find the extreme values in the dataset.
    \item $t_{anomaly}$: The adaptive anomaly threshold.
    \item $\gamma$ and $\sigma$: The shape and scale parameters of the GPD.
    \item $X_i$: Random variable representing the current data point.
    \item $Y_i=X_i-t \mid X_i>t$: Random variable representing the excess over the threshold.
    \item $N$: The total number of data points in the set $X$.
    \item $N_t$: The total number of extreme values in the set $Y$.
    \item $q$: The risk coefficient, or the desired probability, for the anomalies.
\end{itemize}



The tail of the data distribution is represented by the following function:

\begin{equation}
\overline{F}_t(x)=\mathcal{P}(X-t>x \mid X>t)\underset{}{\sim}\left(1+\frac{\gamma x}{\sigma}\right)^{-\frac{1}{\gamma}}
\label{GPD}
\end{equation}

The shape and scale parameters, $\gamma$ and $\sigma$, are the parameters of the distribution of the extreme values and can be estimated using several methods, e.g., Maximum Likelihood Estimation (MLE), Probability Weighted Moments (PWM). However, the Method of Moments (MOM) is selected for parameter estimation because of its simplicity~\cite{lin2024hierarchical} and its superior performance (in terms of output/results) in the case of time series data (compared to MLE)~\cite{li2021fluxev}. The purpose of MOM is to use sample moments to estimate the unknown parameters of the probability distribution. The mean and variance of the GPD are then computed as follows:



\begin{equation}
\mu=\frac{\sigma}{1-\gamma}\simeq\sum_{i=1}^{N_t}\frac{Y_i}{N_t}\text{, for }\gamma < 1
\label{mean}
\end{equation}

\begin{equation}
S^2=\frac{\sigma^2}{(1-\gamma)^2(1-2\gamma)}\simeq\sum_{i=1}^{N_t}\frac{(Y_i-\mu)^2}{N_t-1}\text{, for }\gamma < \frac{1}{2}
\label{variance}
\end{equation}

With the mean and the variance of the extreme values in the dataset, the estimation of $\gamma$ and $\sigma$ is:

\begin{equation}
\hat{\gamma}=\frac{1}{2}\left(1-\frac{\mu^2}{S^2}\right), \quad \hat{\sigma}=\frac{\mu}{2}\left(1+\frac{\mu^2}{S^2}\right)
\label{gamma}
\end{equation}

\noindent Having all the necessary parameters, the adaptive threshold is calculated:

\begin{equation}
t_{anomaly}=t+\frac{\hat{\sigma}}{\hat{\gamma}}\left(\left(\frac{qN}{N_t}\right)^{-\hat{\gamma}}-1\right)
\label{AT}
\end{equation}

After fitting the distribution tail to the GPD, with a given probability $q$, we calculate the anomaly threshold such that $\mathcal{P}(X>t_{anomaly})<q$~\cite{siffer2017anomaly}. When using the POT method, the higher the initial threshold $t$, the less bias in the extreme events. On the other hand, if $t$ is too high, there will not be enough events in the set of extreme values. Besides, the risk coefficient $q$ should be in the range of $10^{-5}$ to $10^{-3}$ to have a high True Positive (TP) rate, while the FP rate stays low. It is important to note that $t<t_{anomaly}$, so the probability associated with $t$ has to be lower than $1-q$, i.e., $\mathcal{P}(X<t)<1-q$.
In the case of time series data, this operation is performed once at each timestamp, and the anomaly threshold is updated continuously.

\section{RRC Signaling Storms}
In the following, we introduce the RRC signaling storm threat and model its impact, capturing gNB behavior under both attack and high-load events. 

\subsection{Threat Model}\label{threat_model}
RRC signaling storms can create an overload state at the gNB by targeting the control plane's availability. As a result, the gNB runs out of RRC resources, and new UEs who want to join the network will not be able to connect, i.e., transit to the RRC\_CONECTED state~\cite{tabiban2023signaling}.
During a signaling storm attack, one or more MUEs would send multiple Msg3s repeatedly and never complete the RRC procedure with Msg5s.
With a high enough attack rate, the MUE can reserve all the available resources before the gNB releases previous ones, which leads to the unavailability at the gNB~\cite{zhang2024mitigating}. An attacker can use random values for the UE's identity parameter in Msg3, the gNB then sees the Msg3s as coming from different UEs. In addition, the establishment cause can be set as emergency/high-priority access, so that the gNB has to prioritize these Msg3s; more details on this threat model can be found in~\cite{nguyen2025rrc}.

\subsection{Theoretical Model}
\label{theoretical_model}
Our previous research on RRC signaling storms presents a theoretical model that explains the behavior of the gNB under attack/high-load situations~\cite{nguyen2025rrc}. 

\textbf{\textit{Notations:}}
The following notations are used for the RRC signaling storm theoretical model.

\begin{itemize}
    \item $T_W$ -- \textit{Waiting time}: When the gNB receives a Msg3, an RRC resource is reserved for a duration of time, i.e., the waiting time. The gNB will release the resource when the waiting time expires if the gNB does not receive any Msg5.
    \item $N_{max}$ -- \textit{Maximum number of UEs}: The largest number of UEs that the gNB can connect simultaneously, i.e., the number of RRC resources at the gNB.
    \item $N_{BUE}$ -- \textit{Number of connected UEs}: The number of UEs currently connected with the gNB, i.e., the UEs that completed the RRC connection establishment procedure.
    \item $R_{att}$ -- \textit{Attack rate}: The rate of incoming Msg3s sent by the attacker.
    \item $R_{BUE}$ -- \textit{Rate of BUEs}: The rate of incoming Msg3s from legitimate BUEs.
    \item $T_A$ -- \textit{Duration of accept}: The duration in which the gNB is still available and can send Msg4s to respond to new Msg3s.
    \item $T_R$ -- \textit{Duration of reject}: The duration in which all resources are reserved/being used, the gNB is blocked and cannot send out Msg4s.
\end{itemize}


The \textit{duration of accept} ($T_A$) and \textit{duration of reject} ($T_R$) at the gNB are calculated as shown in \eqref{T_A}.

\begin{equation}
T_A=\frac{N_{max}-N_{BUE}}{R_{att}+R_{BUE}}, \quad T_R=T_W-T_A
\label{T_A}
\end{equation}



From the durations of accept and reject, the overall availability rate of the gNB can be calculated as follows:

\begin{equation}
R_{avai}= \frac{T_A}{T_A+T_R}
\label{R_avai}
\end{equation}

In addition to RRC signaling storm attacks, legitimate high-load cases, where the large amount of Msg3s comes from BUEs, are taken into account. The duration of accept still follows the same rule as in the attack case. However, the duration of reject does not end at the end of the waiting time. The duration of reject only ends, i.e., the gNB is available again, when one or more UEs disconnect from the network and free the resources they used.

\section{System Design}
\label{Methodology}
The proposed detection system has two steps, as shown in Fig.~\ref{fig:system_diagram}. Step 1 is anomaly detection, where the system detects if there is an abnormal event in the traffic. When an abnormal event is detected, the anomaly detector raises an alert to Step 2, the differentiator. The differentiator considers more data to distinguish between an attack and a legitimate high-load case. By having the second step, the system can know the nature of the abnormal behavior and can then perform mitigation actions accordingly. A high-load should not be treated in the same way as a malicious attack, i.e., legitimate connection requests should not be rejected.

{
\setlength{\belowcaptionskip}{-15pt}
\begin{figure*}[htbp]
\centerline{\includegraphics[width=\linewidth]{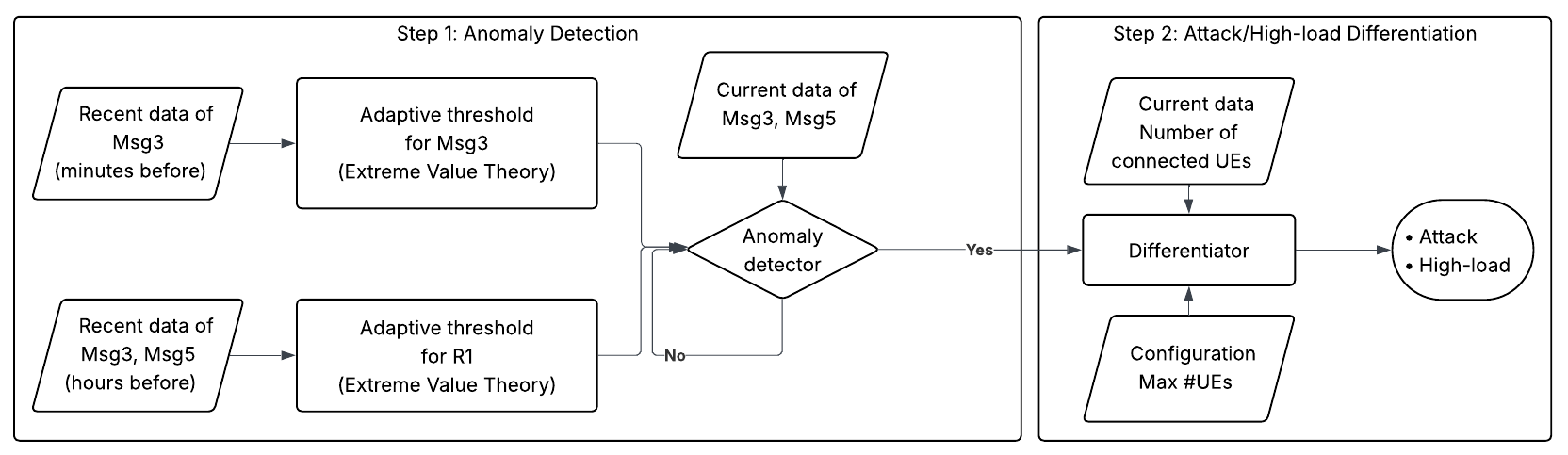}}
\caption{Proposed RRC Signaling Storm Detection System.}
\label{fig:system_diagram}
\end{figure*}
}

\subsection{Anomaly Detection}
\label{sec:anomaly_detection}
The anomaly detector is based on two adaptive thresholds (namely $\#Msg3$ and $R1$ -- explained in the following) that are calculated using RRC features.
Having the detector use two adaptive thresholds for two features improves the detection results since the two features help each other to eliminate FPs. \textbf{When both features exceed the threshold values, the anomaly detector raises alert and trigger the attack/high-load differentiation}. Otherwise, this detector continuously monitors the data.

The EVT method is selected for calculating two adaptive thresholds for the number of Msg3s and the ratio $R1$. By utilizing the EVT, the adaptive thresholds only model the behavior of the extreme tails of the data distribution, i.e., the very rare events; this is where the abnormal events occur. In addition, the EVT also takes into account concept drifts when the data change due to the change in traffic load. To do this, the time window of input data should contain enough extreme data points for the calculation, but also should not be too large to avoid irrelevant fluctuation in the dataset. This results in a suitable threshold value for different times of the day, different cells, and different conditions.

\subsubsection{$\#Msg3$}
This is the number of Msg3s that arrive at the gNB. The number of Msg3s is measured every second and can directly reflect an anomaly when this feature increases significantly more than usual traffic.

On a normal weekday, the traffic load at a gNB follows a specific pattern. The number of Msg3s at the gNB is high during the peak hours, i.e., in the morning and evening. On the other hand, $\#Msg3$ is lower at night.
Because of this, it is not effective to have a static threshold for $\#Msg3$ throughout the day.
Due to the rapid change in the number of incoming Msg3s, recent data from a few minutes before the current time is chosen to be used for the calculation of an adaptive threshold using EVT. The threshold for $\#Msg3$ is called $TH_{Msg3}$, and is calculated with EVT as explained in section~\ref{EVT}.

The majority of the traffic data falls below this threshold. But sometimes, there are still some FPs, i.e., when some normal data points are above the threshold. These data points will be verified using an adaptive threshold for the $R1$ values.
    
\subsubsection{$R1$}
Only looking at $\#Msg3$ may not be enough to detect an abnormal event, then the ratio $R1$ is proposed as a supporting feature to help reduce FPs and improve the detection results.

\begin{equation}
R1=\frac{\#Msg5}{\#Msg3}
\label{R1}
\end{equation}

This ratio has different values between a normal traffic load and when an attack/high-load occurs. In a normal situation, the legitimate UEs can connect to the gNB,
hence,
the number of Msg5s that the gNB receives should be approximately close to the number of Msg3s. The ratio $R1$ has values that are approximately close to $1$ in this case.
However, during an attack/high-load case, the number of Msg5s will decrease while the number of Msg3s remains high.
In fact, during an attack, the attacker does not complete the RRC procedure with Msg5s and continues to send as many Msg3s as possible. Meanwhile, due to the overload state that the gNB experiences in a high-load scenario, the number of Msg5 also decreases.
As a result, the value of $R1$ will decrease towards $0$~\cite{nguyen2025rrc}.


The input to this method is the recent data
of the number of Msg3s and Msg5s.
Different from $\#Msg3$, $R1=1$ most of the time when the network operates in normal conditions, i.e., without extreme events or attacks. Moreover, the value of $R1$ only ranges from 0 to 1, and does not change rapidly as $\#Msg3$. Thus, to obtain enough data for the EVT, the adaptive threshold is computed using a moving window of data spanning a few hours before the current time. 

In the case of $R1$, the extreme values are considered to be the values that are smaller than the initial threshold.
The tail distribution function \eqref{GPD} is modified to consider the differences between the extreme low values and the initial threshold, $Y_i=t-X_i \mid X_i<t$. In this case, \eqref{TH_R1} replaces \eqref{AT} as the adaptive anomaly threshold of $R1$, so that $\mathcal{P}(X<TH_{R1})<q$. In addition, the risk coefficient $q$ is selected as explained in section \ref{EVT}, and the initial threshold $t$ is chosen as a low quantile of the data window, ensuring that $\mathcal{P}(X>t)<1-q$.

\begin{equation}
TH_{R1}=t-\frac{\hat{\sigma}}{\hat{\gamma}}\left(\left(\frac{qN}{N_t}\right)^{-\hat{\gamma}}-1\right)
\label{TH_R1}
\end{equation}

\subsubsection{Algorithm}
Algorithm \ref{TH_alg} shows the core steps of the algorithm for computing the adaptive threshold for $\#Msg3$ and the ratio $R1$ with the application of the POT method.
Because extreme values are considered differently for $\#Msg3$ and $R1$, the calculation of the excess over the initial threshold $t$ is the only difference for them, as shown in line~\ref{line:excess}.


\begin{algorithm}
\caption{Adaptive Threshold Calculation}
\label{TH_alg}
\begin{algorithmic}[1]
\REQUIRE traffic data ($\#Msg3$, $\#Msg5$), risk coefficient $q$,\\time window $N$
\ENSURE initial threshold $t$, adaptive threshold $TH$
\STATE $t \gets InitialThresholdCalculation(\#Msg3, \#Msg5)$
\STATE $Y_{Msg3} \gets \{\#Msg3-t \mid \#Msg3>t\}$ \\ $Y_{R1} \gets \{t-R1 \mid R1<t\}$ \label{line:excess}
\STATE $\hat{\gamma}, \hat{\sigma} \gets MOM(Y)$
\STATE $TH \gets ThresholdCalculation(q,\hat{\gamma},\hat{\sigma},N,N_t,t)$
\end{algorithmic}
\end{algorithm}

The two initial thresholds, and two anomaly thresholds $TH_{Msg3}$, $TH_{R1}$ are calculated based on recent data points. The value of $\#Msg3$ and $R1$ are then compared with these thresholds to decide whether this is an extreme value or not.
These comparisons result in three possible outcomes.

\begin{itemize}
    \item If the current data point exceeds the anomaly threshold, it is identified as an anomaly, and will not be used for the calculation of future thresholds. The anomaly detector sends a signal to the differentiator.
    \item If the current data point is in between the anomaly and the initial threshold, it is identified as an extreme event but not an anomaly, and will be used to calculate future threshold values using the POT method.
    \item If the current data point does not exceed the initial threshold, it is considered a normal event; the anomaly threshold remains the same.
\end{itemize}


\subsection{Attack/High-load Differentiation}
\label{Differentiator}
When an abnormal event occurs, the anomaly detector alerts the attack/high-load differentiator to further analyze the event and check if it results from an attack or a legitimate high-load case. The differentiation step requires two features as input, i.e., the maximum number of UEs that can reach the connected state ($N_{max}$) and the number of currently connected UEs ($N_{BUE}$), and computes their ratio $R2$.

\begin{equation}
R2=\frac{N_{BUE}}{N_{max}}
\label{R2}
\end{equation}

Consistently with the threat model introduced in~\ref{threat_model}, during an attack, even though there are many incoming Msg3s, the number of connected UEs is not expected to increase and will remain smaller than the maximum number of UEs can be connected to the gNB at any time.
Thus, $R2$ can only decrease when connected BUEs disconnect from the gNB.
In contrast, in a high-load case, when the gNB goes to overload state, the number of connected UEs can reach $N_{max}$, because they are legitimate UEs that ultimately want to connect to the network. The value of $R2$ might reach $1$ in the high-load case.

\section{Data Preparation}
\label{data}
\subsection{Legitimate Data}
The proposed detection system is evaluated using real-world telecommunication traces collected from a live commercial LTE/5G network.
The dataset spans across multiple weeks in January 2025, and contains the number of Msg3s, Msg5s, and the average number of currently connected UEs ($N_{BUE}$) at one gNB cell in a busy metropolitan network. 
For our analysis, we extracted data for 4 consecutive weekdays (Tuesday, 2025-01-07 to Friday, 2025-01-10). Because licensing restrictions prevent us from releasing the raw data, we support reproducibility by providing summary statistics (Table~\ref{tab:dataset-stats}) and cumulative distributions (Fig.~\ref{fig:side_by_side}) of collected data. 

\begin{table}[t]
    \centering
    \caption{Statistics of the collected dataset.}
    \begin{tabular}{l r r r}
        \toprule
          & Average & Std.\ dev. & $95^{th}$ percentile\\ 
        \midrule
        \textbf{\#\,Msg3/s}              & 3.14 & 2.73 & 8.546\\[2pt] 
        \textbf{\#\,Msg5/s }             & 3.13 & 2.73 & 8.514\\ [2pt]
        \textbf{Connected UEs}         & 56.51 & 37.87 & 118.548\\  
        \bottomrule
    \end{tabular}
    \label{tab:dataset-stats}
\end{table}

{
\setlength{\belowcaptionskip}{-15pt}
\begin{figure}[!t]
  \centering
  \subfloat[][Connection intensity]{\includegraphics[width=0.48\columnwidth]{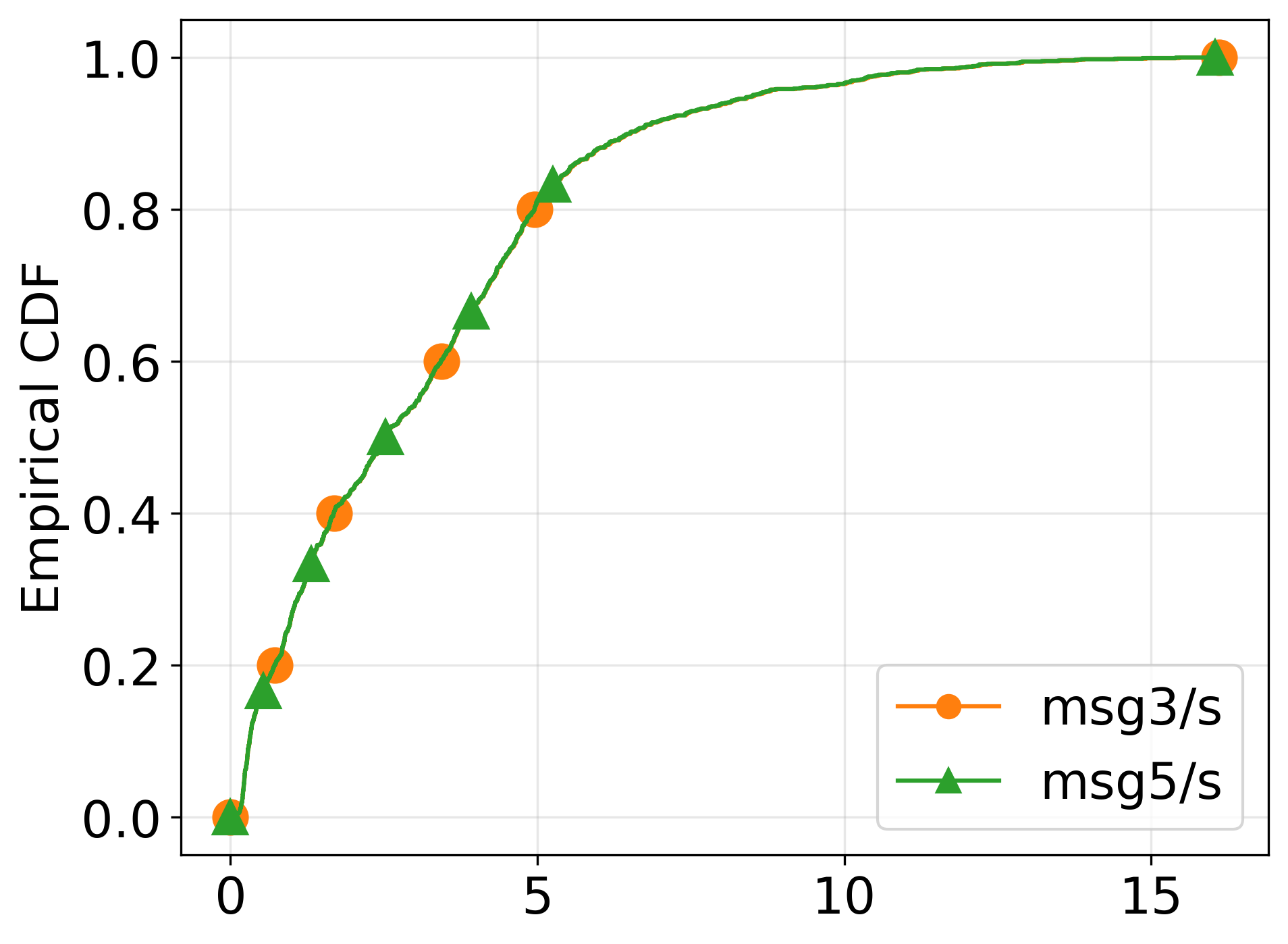}     \label{fig:conn_intens}}
   \subfloat[][Average number of connected UEs]{\includegraphics[width=0.48\columnwidth]{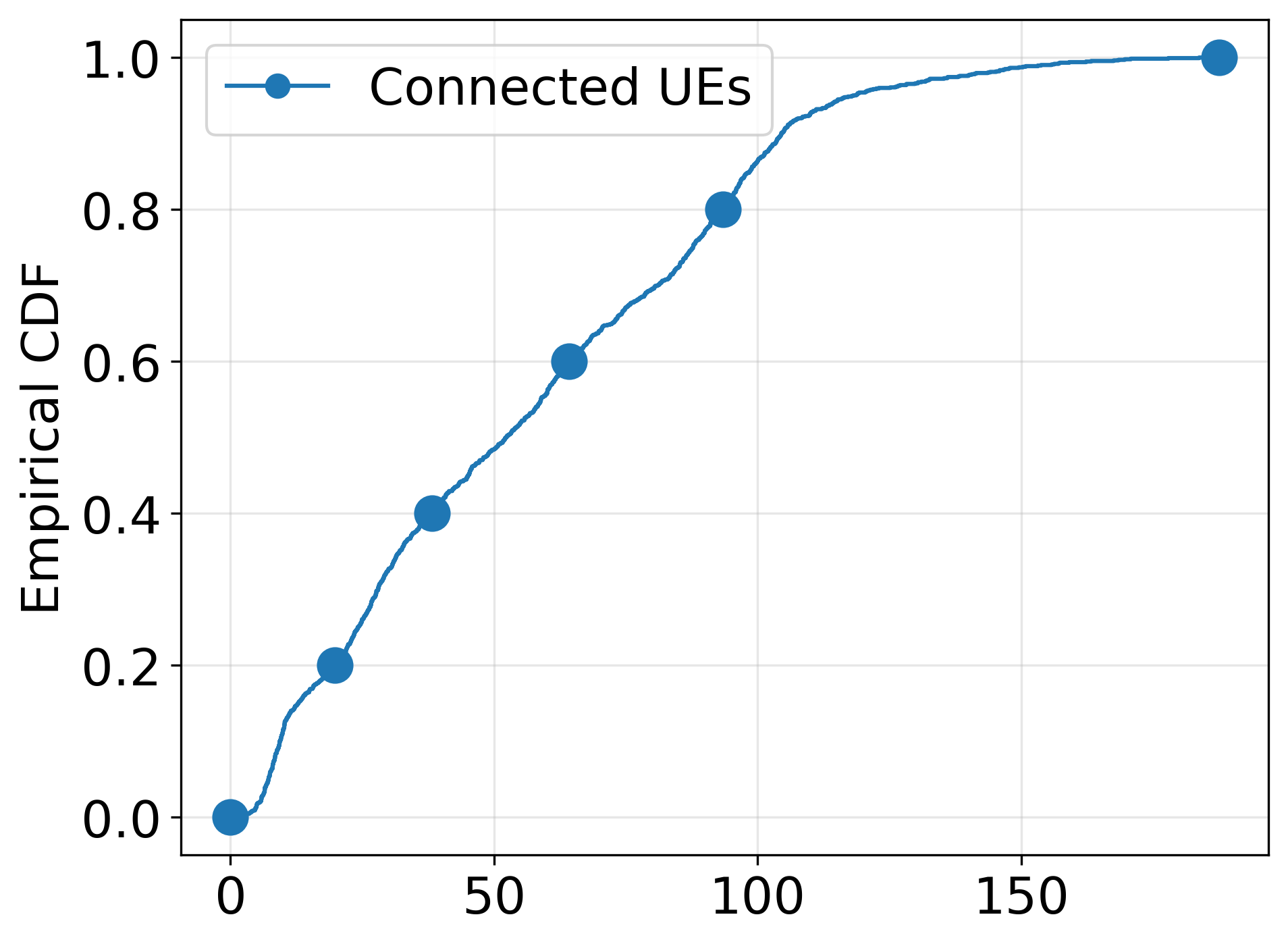}    \label{fig:conn_UEs}}
  \caption{Empirical statistics for the features of interest.}
  \label{fig:side_by_side}
\end{figure}
}

For practicality and scalability, the gNB aggregates data over a 15 minutes period per cell. Since this granularity is not sufficient to perform quick detection in case of an attack, we proceed to resample the data to a granularity of seconds.

\subsubsection{$\#Msg3$}
For the number of Msg3s,
we convert first the values from messages per 15 minutes (i.e., 900 seconds) to messages per second by dividing the recorded values by 900.
Then, the number of Msg3s is resampled using a truncated Poisson distribution. This Poisson distribution has the mean of the actual value of the number of messages per second, and the upper bounds of two times the actual value\footnote{It has to be noted that only in case the mean of a Poisson distribution is large, its distribution becomes approximately symmetric, therefore can be well approximated by a Gaussian distribution.}.

\subsubsection{$\#Msg5$}
A Poisson distribution cannot just be used directly to resample $\#Msg5$ because the number of Msg5s also depends on the number of Msg3s. Hence, it is necessary to look at how they are related to each other. Every Msg5 is preceded by a Msg3, but not all Msg3s are followed by a Msg5. Hence, the relationship between them is shown by the number of failures, i.e., $\#Msg3 - \#Msg5$. From the collected dataset, the number of failures per second is close to 0 and is negligible. Then we can assume that every second, there are either 1 or 0 failed messages. The probability of failure per second and the probability for the value of $\#Msg5$ per second are calculated using \eqref{P_failure} and \eqref{P_Msg5}, respectively.

\begin{equation}
P(failure)=\frac{\#Msg3 - \#Msg5}{900}
\label{P_failure}
\end{equation}

\begin{equation}
P(\#Msg5=\#Msg3-x)=
\begin{cases}
  P(failure) & \mid x=1 \\
  1 - P(failure) & \mid x=0
\end{cases}
\label{P_Msg5}
\end{equation}

\subsubsection{$N_{BUE}$}
The number of connected UEs is used during abnormal events to distinguish between an attack and a high-load. It is therefore not used in the legitimate case.

\subsection{Abnormal Data}
\label{Abnormal_data}
There is no data similar to an RRC signaling storm in the collected dataset. To add abnormal data, we use the theoretical model from Section \ref{theoretical_model}. This theoretical model was verified with experiments in a controlled 5G environment~\cite{nguyen2025rrc}, hence, using it to simulate abnormal events would make attacks and high-load cases realistic. The waiting time $T_W$ is set to 5 seconds. In the recorded data, the highest number of UEs that connect to the gNB at one point in time is around 175 UEs, we make the assumption that this gNB can handle at most 300 UEs simultaneously ($N_{max}=300$). The number of connected UEs ($N_{BUE}$), and the rate of BUEs ($R_{BUE}$) are taken from the dataset.

The highest attack rate is set based on previous work on this topic, i.e., $R_{att\_max}=100$ Msg3s per second~\cite{nguyen2025rrc}. The lowest attack rate is defined as the minimum attack rate that can cause the overload state at the gNB, i.e., $T_R>0$. From \eqref{T_A}, the lowest rate can be computed as follows:

\begin{equation}
R_{att}>R_{att\_min}=\frac{N_{max}-N_{BUE}}{T_W}-R_{BUE}
\label{rate_min}
\end{equation}

To simulate abnormal data, a random attack rate can be chosen from the range of $R_{att\_min}$ to $R_{att\_max}$. This range represents the rate of abnormal activities that can affect the gNB, ranging from minor to severe impact. During the attacks and high-loads, the number of Msg3s is the sum of the number of actual Msg3s and the attack's rate, or the high-load's rate, accordingly.

\begin{equation}
\#Msg3_{att/hl}=\#Msg3+R_{att/hl}
\label{Msg3_att}
\end{equation}

For RRC signaling storm attacks, these five parameters ($T_W$, $N_{max}$, $N_{BUE}$, $R_{BUE}$, and $R_{att}$) are used to calculate the duration of accept ($T_A$) and the duration of reject ($T_R$).
During the duration of accept, the number of Msg5s remains unchanged. On the other hand, during the duration of accept of a high-load, the number of Msg5s is the number of legitimate Msg5s plus the rate of the high-load \eqref{Msg5_hl}. The number of Msg5s is equal to 0 during the duration of reject for both cases.

\begin{equation}
\#Msg5_{hl}=\#Msg5+R_{hl}
\label{Msg5_hl}
\end{equation}

Along with the increase in $\#Msg5$ when the high-load occurs, the number of connected UEs ($N_{BUE}$) also increases until it reaches the capacity of the gNB, i.e., $N_{BUE}=N_{max}$. Because of this, in a simulated attack, the ratio $R2$ stays lower than 1, but increases towards the value of 1 and stays equal to 1 in a high-load.

\section{Implementation \& Evaluation}
\label{implementation_evaluation}
\subsection{Adaptive Threshold Calculation Implementation}

\subsubsection{$\#Msg3$}
The time window for the threshold calculation is set to 3 minutes, as the rate of incoming Msg3s can change rapidly.
Besides, to prevent an attacker from bypassing the adaptive threshold by gradually increasing the attack rate, a 30-second buffer is applied: the most recent 30 seconds of data are excluded from the threshold computation, and only the 3 minutes of data preceding this buffer are used.
In the bootstrap phase, the threshold $t$ is empirically set to a very high quantile, 98\%, and the risk coefficient $q$ is set to the value of $3*10^{-4}$.

\subsubsection{$R1$}
As it can be seen from Fig.~\ref{fig:conn_intens}, the majority of the time, the value of $R1$ is equal to $1$ because in non-overloaded cells there are not many failures in the RRC connection establishment. Then, the time window for the input to the EVT technique needs to be large enough to cover a sufficient amount of extreme events, i.e., low $R1$ values. We found experimentally that a time window set to $5$ hours provides good results for the adaptive threshold $TH_{R1}$. In the bootstrap phase, the value of $t$ is chosen as the $0.1\%$ percentile, and the risk coefficient $q$ is fixed at $10^{-5}$.

In the event of an RRC signaling storm, the value of $R1$ may exhibit a gradual decline rather than a sudden drop.
To prevent the changing trend in the value affecting the adaptive threshold, we implemented a buffer between the input data of the threshold and the current time. From the current data point, there is a 1-minute gap of unused data, and then 5 hours of past data are used to calculate the adaptive threshold.

\subsection{Detection Results \& Evaluation}

{
\setlength{\belowcaptionskip}{-10pt}
\begin{figure}[t]
\centering
\subfloat[][The Number of Msg3s and The Abnormal Threshold]{\includegraphics[width=0.49\textwidth]{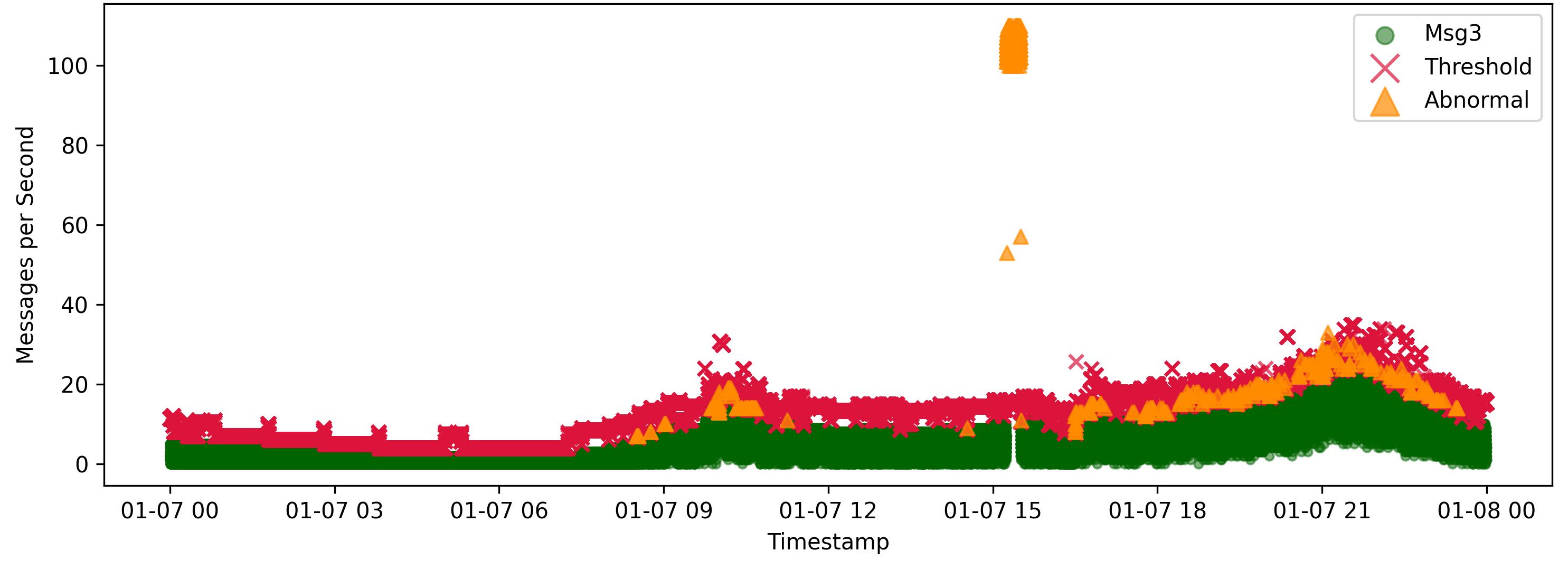}\label{result_Msg3_1attack}}\\
\subfloat[][R1 and The Adaptive Anomaly Threshold]{\includegraphics[width=0.49\textwidth]{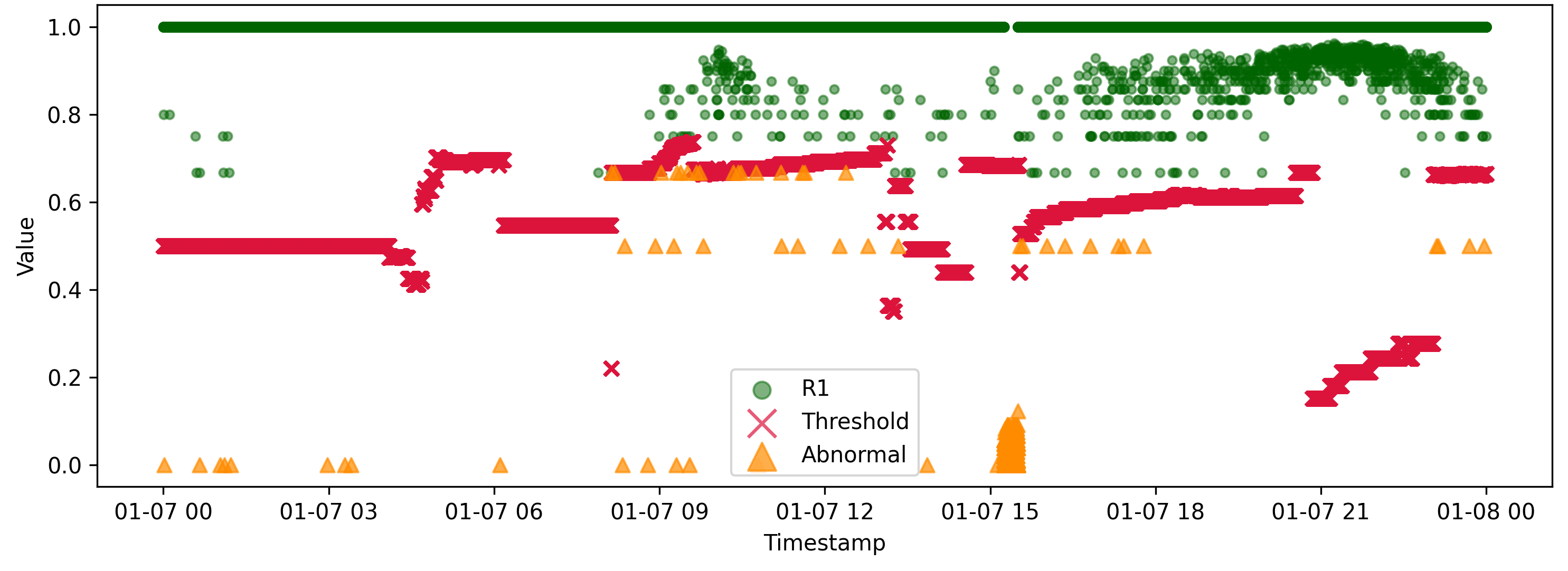}\label{result_R1_1attack}}
\caption{Data plot for 1 day - single attack.}
\label{one_attack}
\end{figure}
}

\subsubsection{Single Attack}
As initial validation scenario, we test the response of our system against a single attack over a 24-hour trace. Beginning at 15:15 on January 7, the adversary floods the gNB for 15 minutes with an attack rate set at $100$~Msg3/s. 

Fig.~\ref{one_attack} shows the performance of the both the thresholds for $\#Msg3$ and $R1$.
One can note that both thresholds effectively adapt to changes in the data without being influenced by abnormal observations. This test has an imbalance between the normal and abnormal data, so only precision and recall are relevant as evaluation metrics. The results for the thresholds for $\#Msg3$ and $R1$ have precision values of $75.08\%$ and $94.45\%$, respectively, and both have recall values of $100\%$. Combining these two results per Section~\ref{sec:anomaly_detection}, there are no FP and no FN, i.e., all normal and attack data are detected correctly.

Although these results are encouraging, the considered scenario is quite simplistic, with the attack traffic clearly distinguishable from the background. 
We now consider how the detection system performs under more challenging conditions, such as multiple attacks/high-loads, different attack/high-load rates, and attacks/high-loads during different times of the day.

{
\setlength{\belowcaptionskip}{-15pt}
\begin{figure}[t]
\centering
\subfloat[][The Number of Msg3s and The Abnormal Threshold]{\includegraphics[width=0.49\textwidth]{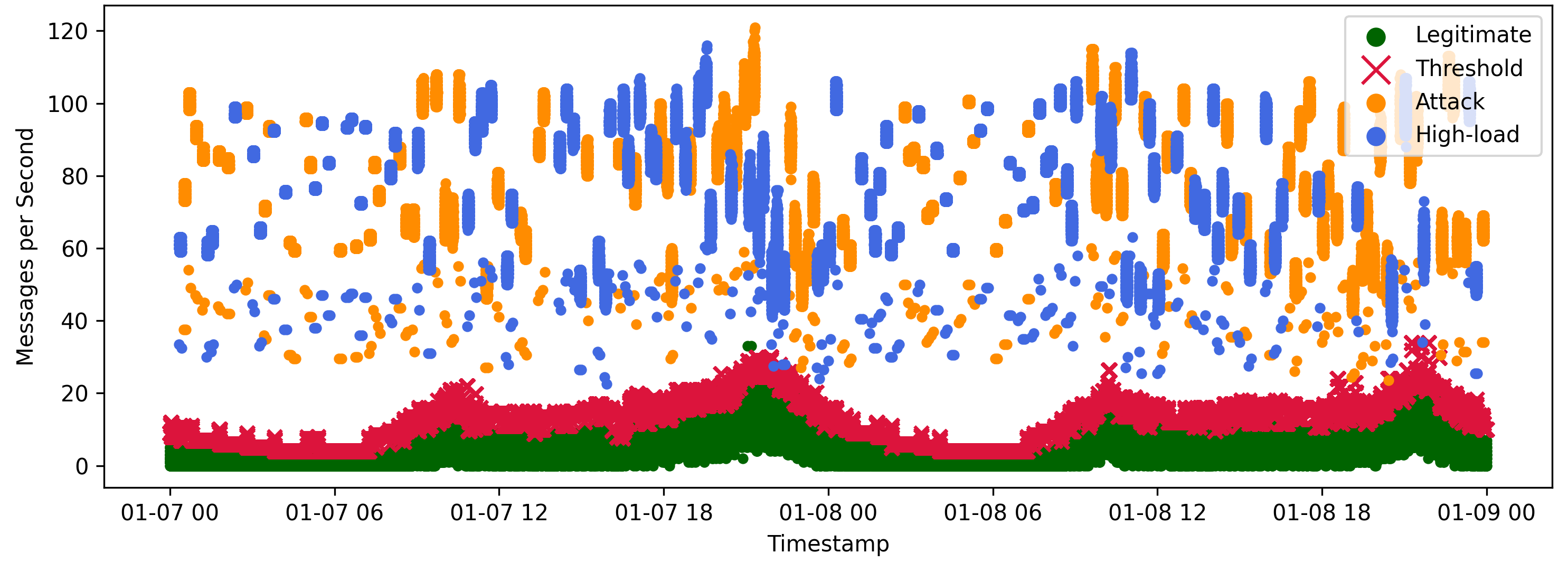}\label{result_original_msgs}}\\
\subfloat[][R1 and The Adaptive Anomaly Threshold]{\includegraphics[width=0.49\textwidth]{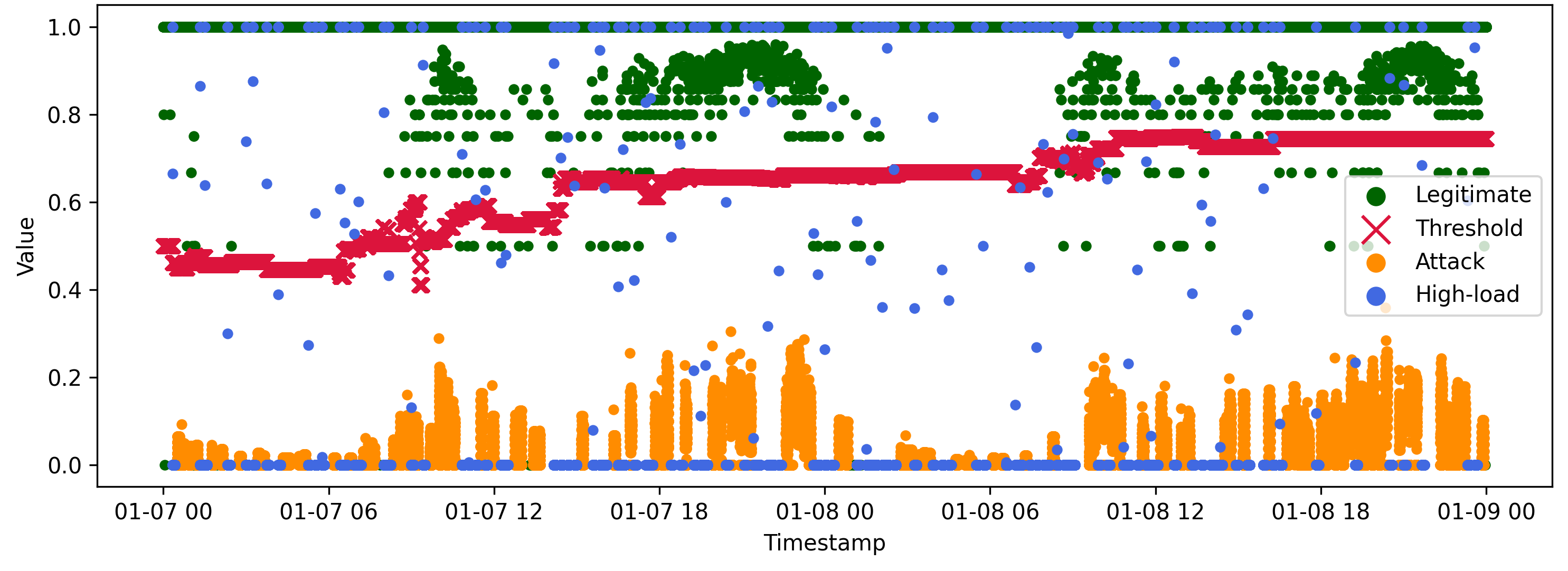}\label{result_original_R1}}
\caption{Data plot for 2 days including legitimate (green), attack (yellow), and high-load (blue) periods.}
\label{result_original}
\end{figure}
}

\subsubsection{Multiple Attacks/High-loads}
\label{multiple_attacks}
To fully evaluate the proposed detection system, we consider the entire 4 weekdays trace, and slice it into 5-minute periods, ($1152$ periods in total). Each period is randomly labeled as attack, high-load, or normal. To avoid trivial long runs of anomalies, attack and high-load periods are inserted only between two normal periods: multiple consecutive anomalous periods would simply prolong the event without affecting the detection performance.

For an attack/high-load, the rate is chosen from the range between $R_{att\_min}$ to $R_{att\_max}$ introduced in Section~\ref{Abnormal_data}.
To suppress oscillations, the detector uses a short confirmation buffer. Successful detection of an attack/high-load is considered if abnormal behavior is raised/flagged for 2 or more consecutive positives.
Fig.~\ref{result_original} shows the detection results over 2 days for clearer visualization, although the evaluation was conducted over the full 4-day duration, with consistent findings throughout.
In total, there are $737$ normal periods, $211$ attacks, and $204$ high-loads. All of them were successfully detected without any FP or FN. Moreover, on average, the attacks/high-loads are detected only within $2.72 s$, which can help the network perform the necessary mitigation technique quickly to ensure the quality of the service.

\subsubsection{EVT vs. Gaussian model}
One key question is how the proposed EVT-based system performs against a simpler Gaussian-threshold adaptation method, which assumes that the number of Msg3 arrivals per second is normally distributed (as mentioned above, this is a good approximation only when the mean is sufficiently large). 

Due to the similarity of traffic pattern in different weekdays, it is possible to use the historical data of a previous weekday as a baseline to know what traffic pattern should be expected, i.e., using the traffic on Monday as a baseline to calculate the anomaly threshold for the following weekdays.
According to the Empirical Rule for a Gaussian distribution, 99.7\% of the data points should be included within 3 standard deviations ($3\sigma$) from the mean ($\mu$)~\cite{prasad2022elementary}. This threshold can be applied for $\#Msg3$ as a baseline method. In that case, the day is divided into 2-hour periods during which the number of incoming Msg3s remains relatively stable, despite fluctuations in traffic throughout the day. The anomaly threshold is then calculated using this ``$3\sigma$" rule for each period.

Table~\ref{tab:summary} reports detection results on the same validation set from the previous test case. The EVT-driven threshold increases Msg3s precision from $72.61\%$ to $95.67\%$, over the Gaussian baseline (R1 shows similar improvement). We attribute this improvement to the fact that the Gaussian baseline tends to systematically underestimate tail probabilities in heavy tail distributions (meaning less TPs). On the contrary, EVT, whose threshold is calculated on the empirical tail, correctly identify benign events while flagging true attacks.

\subsubsection{Attacks with low unavailability rate}
Fig.~\ref{result_original_msgs} shows a clear separation between legitimate and abnormal traffic. A detector that simply tracks $Msg3s$ counts could already perform very well. Consequently, we have defined more challenging test scenarios to further assess the performance of our approach.
For instance, we simulate an adversary that generates low-intensity attacks that degrades, but not fully halt, gNB service. The attack rate is set via~\eqref{R_avai} to yield 5\% unavailability rate ($R_{avai}=95\%$). Fig.~\ref{results_unav_rate}  shows the test data and resulting adaptive threshold values. Table~\ref{tab:summary} reports performance results. Despite a few additional FNs where the detection system misses attacks, the overall performance remains strong, with accuracy of 97.1\%, precision of 100\%, and recall of 93.4\%. We omit high-load events here because when they occur, the gNB is expected to remain in an overloaded state for an extended interval, making this scenario unrealistic. 

\subsubsection{Attacks/High-loads with low attack rate} Next, we consider even more subtle attacks where the gNB does not reach the overload state. Fig.~\ref{result_low_rate} depicts the test data and resulting EVT threshold when the attacker transmits between 50\% to 100\% of the minimum overload rate $R_{att\_min}$. In this case, the attacker fails its objective of blocking BUEs connections, yet manages to increase the processing resources at the gNB. Since the attack we want to identify is often well inside the boundaries of ordinary traffic bursts, this is a challenging scenario for the detector. From the performance results reported in Table~\ref{tab:summary}, we can see that FPs remain low, confirming the performance of the EVT method, even after the tail has moved closer to the bulk of the distribution. Precision remains at 100\%, guaranteeing that every detected event is an attack. A few isolated FNs drive down the recall (96.48\%), explained by the fact that in some cases, the attack merges into the background traffic.

\subsubsection{Busy gNB} Finally, RRC signaling storms may occur at an already busy site. In this case, it is easy for an attacker to flood a gNB that is already under high traffic load, making at the same time more difficult to differentiate abnormal behavior. From the original legitimate data, we simulate this scenario by scaling up the traffic data by 1.5 times. We add abnormal data in the same way as in Section~\ref{multiple_attacks}. Resulting traffic and threshold is plotted in Fig.~\ref{result_high_traffic}.


{
\setlength{\belowcaptionskip}{-15pt}
\begin{figure}[t]
\centering
\subfloat[][Attacks with low unavailability rate.]{\includegraphics[width=0.49\textwidth]{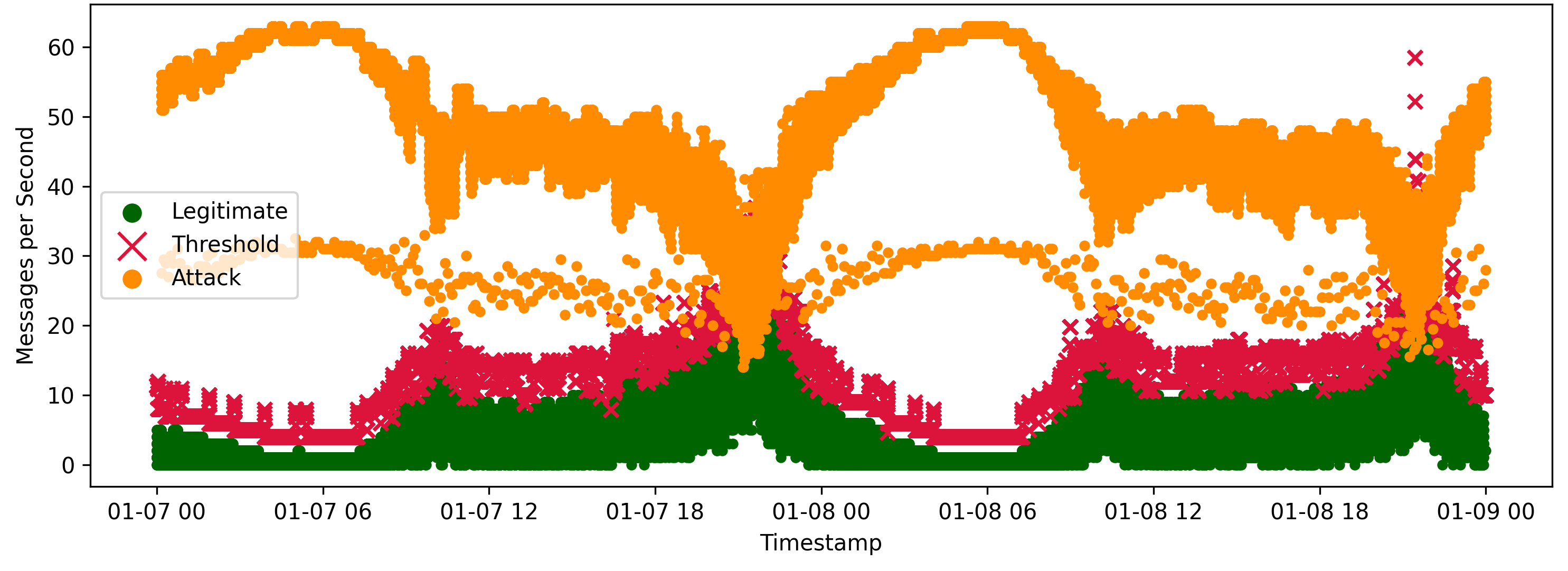}\label{results_unav_rate}}\\
\subfloat[][Attacks/High-loads with low attack rate.]{\includegraphics[width=0.49\textwidth]{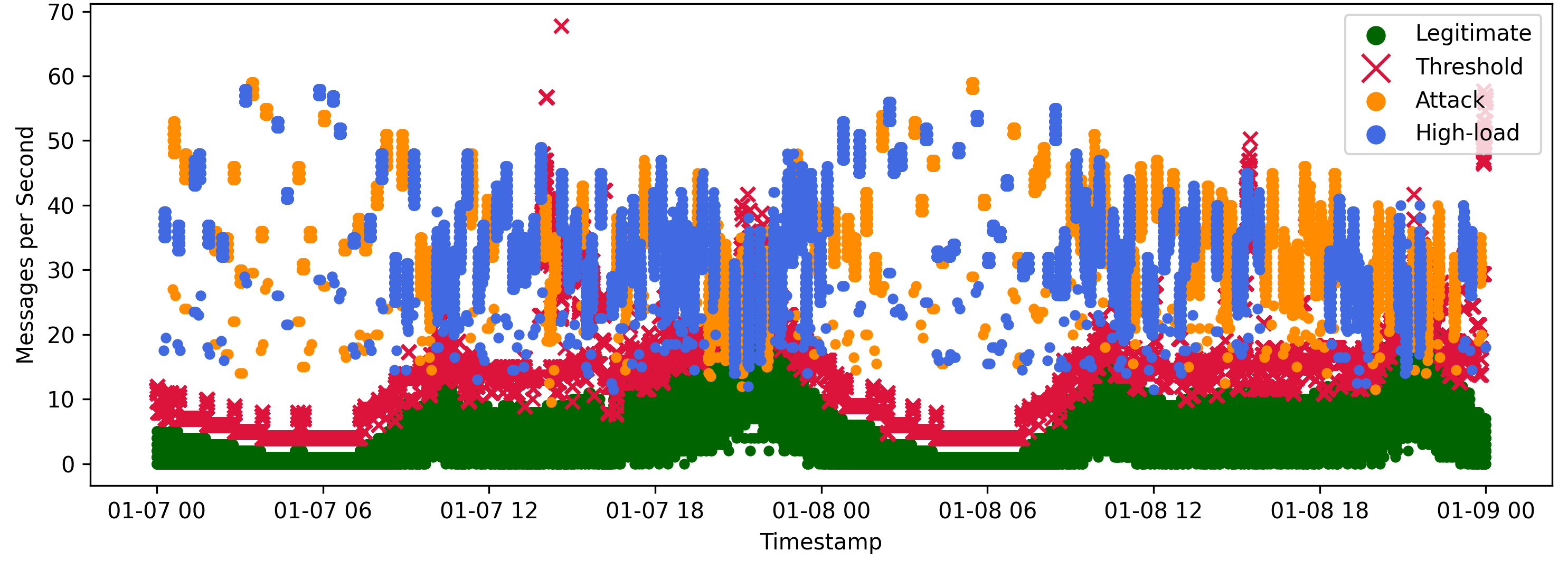}\label{result_low_rate}}\\
\subfloat[][Attacks/High-loads targeting a busy gNB.]{\includegraphics[width=0.49\textwidth]{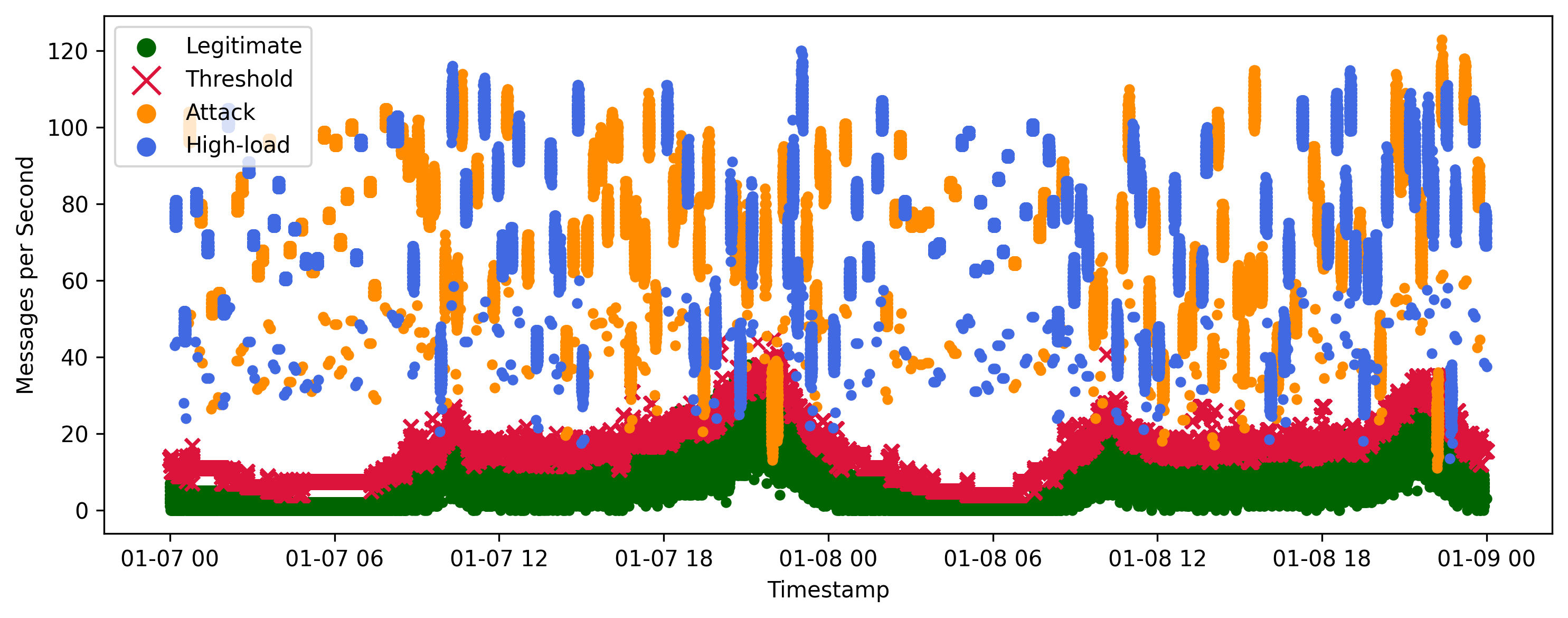}\label{result_high_traffic}}
\caption{The Number of Msg3s and The Abnormal Threshold for Complex Evaluation Scenarios.}
\label{result_complex}
\end{figure}
}

\begin{table*}[!htbp]
\caption{Evaluation of The Proposed Detection System.}
\label{tab:summary}
\begin{center}
\begin{tabular}{|c|l|c|c|c|c|c|c|} 
 \hline
 \multicolumn{1}{|c|}{\textbf{Category}} & \multicolumn{1}{|c|}{\textbf{Scenario}} & \multicolumn{1}{|c|}{\textbf{Method}} & \multicolumn{1}{|c|}{\textbf{Feature}} & \multicolumn{1}{|c|}{\textbf{Accuracy}} & \multicolumn{1}{|c|}{\textbf{Precision}} & \multicolumn{1}{|c|}{\textbf{Recall}} & \multicolumn{1}{|c|}{\textbf{Latency}}\\
 \hline \hline
 \multirow{3}{*}{Baseline} & \multirow{3}{*}{Multiple random attacks/high-loads} & $\mu+3\sigma$ & $\#Msg3$ & 86.37\% & 72.61\% & 100\% & 2.63s \\
 \cline{3-8}
 &  & EVT & $\#Msg3$ & 98.35\% & 95.67\% & 100\% & 2.72s \\
 \cline{3-8}
 &  & EVT & $R1$ & 99.57\% & 98.82\% & 100\% & 2.63s \\
 \hline \hline
 \multirow{3}{*}{\shortstack{Our\\proposed}} & Multiple random attacks/high-loads & EVT & $\#Msg3$ + $R1$ & 100\% & 100\% & 100\% & 2.72s \\
 \cline{2-8}
 & Attacks with low unavailability rate (Fig.~\ref{results_unav_rate}) & EVT & $\#Msg3$ + $R1$ & 97.1\% & 100\% & 93.4\% & 5.12s \\
 \cline{2-8}
 & Attacks/High-loads with low attack rate (Fig.~\ref{result_low_rate}) & EVT & $\#Msg3$ + $R1$ & 98.7\% & 100\% & 96.48\% & 5.81s \\
 \cline{2-8}
 & Attacks/High-loads targeting a busy gNB (Fig.~\ref{result_high_traffic}) & EVT & $\#Msg3$ + $R1$ & 99.65\% & 100\% & 99.06\% & 2.71s \\
 \hline
\end{tabular}
\end{center}
\end{table*}

Like in the other two complex test cases, normal and abnormal data are close to each other, sometimes even overlapping. Nevertheless, the proposed detection system still performs well and can detect anomalies and distinguish between attack and high-load situations. The comparison with other methods, and the summary of all evaluation scenarios are shown in Table~\ref{tab:summary}.
An interesting aspect for the two test scenarios 4) and 5) is that they stress the detection system. One visible result is that the EVT-based detector needs additional time to consistently flag attacks and high-loads, showing as well slightly lower accuracy and recall than other scenarios. A possible reason is that by relying on POT method coupled with a confirmation buffer requires to identify additional consecutive periods with excess burst over the threshold. Naturally, this takes longer when the attack rate is low. However, in all cases, the precision has the value of $100\%$; no attack is mistaken as a high-load and vice versa. 

\section{Conclusion}
\label{conclusion}
In this paper, we have investigated the RRC signaling storms in mobile networks, which can cause service disruption at base stations and prevent legitimate UEs from connecting to the gNB. To address this challenge, we proposed an adaptive threshold-based detection system that leverages statistical techniques to identify such attacks, and then distinguish them from legitimate high-traffic scenarios. The detection system was evaluated using real benign traffic and simulated attacks derived from a realistic theoretical model.
The validation shows that the system achieved high performance results with low detection latency, even in complicated situations, demonstrating the effectiveness and robustness of the approach.
Unlike previous solutions that rely on static thresholds, our method offers an adaptable alternative to the dynamics of traffic using Extreme Value Theory.
Although results are promising, the absence of real-world abnormal data may limit the generalization of results. Future work will focus on exploring mitigation strategies that can be automatically triggered upon detection to enhance network resilience against RRC signaling storms.

\bibliographystyle{IEEEtran}
\bibliography{ref}

\vspace{12pt}

\end{document}